\documentstyle[seceq,amssymb,graphicx,epsf]{ptptex}

\newcommand{\nw}{\nu(r,t)}
\newcommand{\lam}{\lambda(r,t)}

\newcommand{\C}{8\pi}
\newtheorem{theorem}{Theorem:}

\notypesetlogo  

\title{
Spherical Gravitating Systems of Arbitrary Dimension }

\author{
A {\sc Das}\footnote{E-mail: das@sfu.ca} and A {\sc
DeBenedictis}\footnote{E-mail: adebened@langara.bc.ca} }

\inst{
$^{*}$\it{\small Department of Mathematics, Simon Fraser University,
Burnaby, British Columbia, Canada V5A 1S6}
\\
$^{**}$\it{\small Department of Physics, Langara College, 100 W.
49$^{\mbox{th}}$ ave., Vancouver, British Columbia, Canada V5Y 2Z6}}

\abst{
We study spherically symmetric solutions to the Einstein field
equations under the assumption that the space-time may possess an
arbitrary number of spatial dimensions. The general solution of Synge
is extended to describe systems of any dimension. Arbitrary dimension
analogues of four dimensional solutions are also presented, derived
using the above scheme. Finally, we discuss the requirements for the
existence of Birkhoff's theorems in space-times of arbitrary dimension
with or without matter fields present. Cases are discussed where the
assumptions of the theorem are considerably weakened yet the theorem
still holds. We also discuss where the weakening of certain conditions
may cause the theorem to fail. }

\begin{document}

\maketitle

\section{Introduction}
\qquad Space-times possessing dimension greater than four have been of
much interest at least since the pioneering ideas of Kaluza
\cite{ref:kaluza} and Klein \cite{ref:klein} \cite{ref:klein2}. Since
then there have been numerous theories of unification (for example
superstring theory), many of which require more than three spatial
dimensions to be consistent. The low energy sector of many of these
theories reduce to a multi-dimensional General Relativity theory as is
studied here.

\qquad Although extra dimensions are usually thought to be compact,
the extra dimensions may be manifest on scales which are relevant when
studying cosmological systems near the big bang or gravitational
collapse approaching the singularity. Also, there has lately been much
interest in the possibility of large extra dimensions
\cite{ref:arkani}$^{-}$\cite{ref:arkani3} in which higher dimensional
effects may be observed at relatively low TeV scales. Microscopic
systems above these scales may be formed from the collapse of large,
effectively four dimensional, initial conditions. If these scenarios
do indeed describe our universe, then gravitating systems at these
scales will behave as higher dimensional systems and deviate from the
predictions of four dimensional physics.

\qquad Much interesting work has been done in the field of higher
dimensional gravity (for example, see
\cite{ref:bron}$^{-}$\cite{ref:kimmoonyee} and references therein. The
excellent review by Melnikov \cite{ref:melrev} has an extensive list
of references) . Usually, these studies involve specific matter fields
or a class of metrics such as the FRW cosmological metrics. However,
little work has been done on a reasonably general methodology which
may be utilised in studying higher dimensional gravitation problems.
Granted, a completely general method to solve Einstein's field
equations does not exist and therefore in this note we focus on
spherically symmetric systems. Since some interesting studies have
been performed on the subject of higher dimensional black holes (for
examples see \cite{ref:myers}, \cite{ref:taylorrob} and references
therein) our concentration here will be on non-vacuum systems.

\qquad Finally, studies of General Relativity in arbitrary dimension
will also serve to shed light on the theory's internal consistencies
and lead to a greater understanding of the theory as a whole. This
avenue has already proved fruitful in the case of low dimensional
black holes \cite{ref:bhtz}$^{-}$\cite{ref:jeongwon}, for example.
\footnote{The literature on lower dimensional black holes is
extensive. We apologize that we cannot cite all the excellent work in
this area.}

\qquad We therefore believe that it is of much interest to study how
the presence of an arbitrary number of spatial dimensions affects the
solutions of general relativity. To this end, in section 2, we
generalise Synge's \cite{ref:syngebook} method of solving the
spherically symmetric Einstein field equations so that it may apply to
$D$-dimensional systems. In section 3 we present solutions which are
arbitrary dimensional counterparts to some important 4 dimensional
solutions. Finally, in section 4, we rigorously prove a $D$-dimensional
staticity or Birkhoff's theorem and comment on situations when the
theorem can fail.

\section{The general solution}
\qquad The fundamental equations governing the space-time geometry may
be derived from the action \footnote{Conventions in this paper follow
those of \cite{ref:MTW} with $G_{D}$, the D-dimensional Newton's
constant, and $c \equiv 1$. Here, Greek indices take on values $0
\rightarrow D-1$ whereas Latin indices take on values $1 \rightarrow
D-1$ (spatial indices).}
\begin{equation}
S=\int\, \left(-\frac{R}{16\pi} +{\mathcal{L}}_{m} \right) \sqrt{g}\,
d^{D}x \;, \label{eq:action}
\end{equation}
where ${\mathcal{L}}_{m}$ is the matter Lagrangian density and the
positive integer $D$ is assumed to be larger than 2.

\qquad The action principle with suitable boundary conditions gives
rise to the $D$-dimensional Einstein field equations:
\begin{equation}
R^{\mu}_{\;\nu}-\frac{1}{2}R\, \delta^{\mu}_{\; \nu}= 8\pi T^{\mu}_{\;
\nu}\; , \label{eq:einst}
\end{equation}
along with supplementary equations governing the behaviour of the
matter fields. We wish to study solutions under the ansatz of
spherical symmetry. In curvature coordinates this allows us to write
the space-time metric as:
\begin{equation}
ds^{2}= -e^{\nw}\,dt^{2}+ e^{\lam}\, dr^{2} + r^{2}\,
d\Omega^{2}_{(D-2)}, \label{eq:spheremetric}
\end{equation}
where $d\Omega^{2}_{(D-2)}$ is the line element on a unit $D-2$ sphere:
\begin{equation}
d\Omega^{2}_{(D-2)}=\left[d\theta_{(0)}^{2}+\sum_{n=1}^{D-3}d\theta_{(n)}^{2}
\left(\prod_{m=1}^{n}\sin^{2}\theta_{(m-1)}\right)\right]. \nonumber
\end{equation}
The corresponding hyper-surface volume is given by:
\begin{equation}
V_{(D-2)-\mbox{sphere}}=\frac{2\pi^{(D-1)/2}}{\Gamma\left(\frac{D-1}{2}\right)}\;\;
, \label{eq:spherevol}
\end{equation}
and the coordinate ranges are:
\begin{equation}
t_{1} < t < t_{2},\; r_{1} < r < r_{2}, \; 0 <
\theta_{(0)},\,\theta_{(1)},\ldots,\theta_{(D-4)} < \pi, \; 0 \leq
\theta_{(D-3)} < 2\pi. \label{eq:ranges}
\end{equation}

\qquad The above metric yields the following field equations for
(\ref{eq:einst}):
\begin{subequations}
\begin{eqnarray}
8\pi T^{t}_{\;t} &=&-\frac{D-2}{2
r^{2}}\left[(D-3)\left(1-e^{-\lam}\right) +r
e^{-\lam} \lam_{,r}\right]\;, \label{eq:einstt} \\
8\pi T^{r}_{\; r} &=& -\frac{D-2}{2
r^{2}}\left[(D-3)\left(1-e^{-\lam}\right) -r
e^{-\lam}\nw_{,r}\right]\;,  \label{eq:einstrr} \\
8\pi T^{t}_{\; r} &=& -\frac{D-2}{2r}e^{-\nw}\lam_{,t}\;,
\label{eq:einsttr} \\
8\pi T^{\theta_{(A)}}_{\; \theta_{(A)}} &=&
\frac{e^{-\nw}}{4}\left[\nw_{,t}\lam_{,t} -
\left(\lam_{,t}\right)^{2} - 2\lam_{,t,t}\right] \nonumber \\
& &\left.+ \frac{e^{-\lam}}{4}\left[2 \nw_{,r,r} +
\left(\nw_{,r}\right)^{2} +
\frac{2(D-3)}{r}\left(\nw-\lam\right)_{,r} \right.\right. \nonumber \\
& &\left. -\nw_{,r}\lam_{,r} +\frac{2}{r^{2}}(D-3)(D-4)\right]
-\frac{2(D-3)(D-4)}{r^{2}} \label{eq:einstthetatheta}.
\end{eqnarray}
\end{subequations}

\qquad Enforcing conservation laws, $T^{\mu}_{\;\nu;\mu}\equiv 0$,
yields the following non-trivial equations:
\begin{subequations}
\begin{eqnarray}
&T^{t}_{\;t,t} + T^{r}_{\;t,r} +\frac{1}{2}\lam_{,t} \left(T^{t}_{\;t}-
T^{r}_{\; r}\right)+\frac{1}{2} T^{r}_{\; t} \left[\left(\lam + \nw
\right)_{,r} + \frac{2(D-2)}{r}\right]
\nonumber \\ &= 0, \label{eq:cons1} \\
&T^{r}_{\;r,r}+T^{t}_{\;r,t}+ \left[\frac{1}{2}\nw_{,r}
+\frac{D-2}{r}\right] T^{r}_{\; r}+
\frac{1}{2}\left[\nw+\lam\right]_{,t} T^{t}_{\; r} \nonumber \\
&-\left[\frac{1}{2}\nw_{,r}T^{t}_{\;t}+
\frac{D-2}{r}T^{\theta_{(A)}}_{\;\theta_{(A)}}\right]=0.
\label{eq:cons2}
\end{eqnarray}
\end{subequations}

\qquad The field equations must now be solved. We adopt here a similar
method to that of Synge \cite{ref:syngebook} which we generalise to
accommodate arbitrary dimension.

\qquad The system possesses six partial differential equations (which
admits two differential identities) and six unknown functions: $\nw$,
$\lam$ and the four relevant components of the stress-energy tensor.
One may therefore either prescribe two of these functions or else they
may be determined by other means, such as supplementary matter
equations. We assume, for the moment, that $T^{t}_{\; t}$ and
$T^{r}_{\; r}$ are the known functions (we will mention more on this
issue shortly). They may also be related by an equation of state. We
solve for $\lam$ via (\ref{eq:einstt}) by noting that (for $D > 2$) it
may be written as:
\begin{equation}
-16\pi\frac{r^{D-2}}{D-2}T^{t}_{\; t}=\left[r^{D-3}w(r,t)\right]_{,r}
\end{equation}
with $w(r,t):=1-e^{-\lam}$. Integrating this equation followed by
minor algebraic manipulation gives:
\begin{equation}
e^{-\lam}=1+\frac{16\pi}{(D-2)r^{D-3}}\int_{r_{0}}^{r}T^{t}_{\;t}(x,t)
x^{D-2}\, dx +\frac{f(t)}{r^{D-3}} =:1-\frac{2m(r,t)}{r^{D-3}} .
\label{eq:lameqn}
\end{equation}
Here $f(t)$ is an arbitrary or free function of integration. To avoid
a singularity at $r=0$ one must set this function to zero. However,
here we shall keep it for generality.

\qquad The metric function, $\nw$, is obtained from a linear
combination of (\ref{eq:einstt}) and (\ref{eq:einstrr}) as:
\begin{equation}
\C\left[T^{t}_{\; t}-T^{r}_{\; r}\right]= -\frac{D-2}{2r} e^{-\lam}
\left[\nw+\lam\right]_{,r} , \label{eq:derivs}
\end{equation}
which, using (\ref{eq:lameqn}), yields
\begin{equation}
e^{\nw}=\left[1- \frac{2m(r,t)}{r^{D-3}}\right]
\exp\left\{h(t)+\frac{16\pi}{D-2}\int_{r_{0}}^{r}
\left[\frac{T^{r}_{\; r}(x,t) - T^{t}_{\;
t}(x,t)}{x^{D-3}-2m(x,t)}\right]x^{D-2} \, dx \right\}.
\label{eq:nueqn}
\end{equation}
The function $h(t)$ is a function of integration which {\em may} be
absorbed in the definition of a new time coordinate via the
transformation $\,\hat{t}=\int\exp\left[h(t)/2\right]dt$. This is not
always possible as will be discussed in section 4.

\qquad The energy flux, $T^{t}_{\; r}$, may now be defined by the
equation (\ref{eq:einsttr}) and the lateral pressure, $T^{\theta}_{\;
\theta}$, is defined from the conservation law (\ref{eq:cons1}):
\begin{eqnarray}
T^{\theta_{(A)}}_{\;\theta_{(A)}}:=&\frac{r}{D-2}\left\{T^{r}_{\;r,r} +
T^{t}_{\; r,t} + \frac{1}{2}\left[\nw+\lam\right]_{,t}T^{t}_{\;r}
\right. \nonumber \\
&+\left. \left[\frac{1}{2}\nw_{,r} +\frac{D-2}{r}\right] T^{r}_{\;r}
-\frac{1}{2}\nw_{,r} T^{t}_{\; t} \right\}, \label{eq:latpres}
\end{eqnarray}
as this must be the lateral pressure if the energy density and
parallel pressure are known. At this point, it can be shown that {\em
all} equations and identities are satisfied.

\qquad If one is interested in solutions given by specific matter
fields, the number of unknowns versus the number of equations may be
different than previously specified. For example, if the system to be
studied respects absolute spatial isotropy (as in the case of a
perfect fluid), then (\ref{eq:latpres}) becomes a differential
equation for $T^{r}_{\; r}=T^{\theta}_{\;\theta}$ which must be
solved. The general solution presented above will still contain these
as specific cases although other constraints must be met (which may be
variationally derived and lead to a determinate system).

\qquad Otherwise, one is free to prescribe the two functions. The most
physical prescription involves specification of the energy density and
one (parallel) pressure. This method is quite useful in examinations of
relativistic stellar structure and collapse dynamics where one usually
prescribes a reasonable energy and pressure from nuclear theory and
studies of plasmas \cite{ref:weinberg}.

\section{D - dimensional counterparts to known solutions}
\qquad In this section we construct D-dimensional analogues of some
four dimensional solutions. It is useful to study specific solutions
since they are illustrative of how various physical properties in the
construction depend on space-time dimension. The first solution we
present is previously known. We briefly present it here to illustrate
how it may quickly be derived using the method of the previous section.

\subsection{Kottler solution}
\qquad Consider the case of a Kottler (Schwarzschild-(anti) de Sitter)
solution \cite{ref:kottler} in $D$-dimensions. In this situation, the
cosmological constant is best viewed as part of $T^{\mu}_{\;\nu}$:
\begin{equation}
T^{\mu}_{\;\nu}=-\frac{(D-2) M}{8\pi
r^{D-2}}\delta(r)\,\delta^{\mu}_{\;t}\delta^{t}_{\;\nu}
-\frac{1}{8\pi}\Lambda\,\delta^{\mu}_{\;\nu},
\end{equation}
$M$ being a constant related to the effective mass of the black hole.
Here we are {\em loosely} using the coordinates $r$ and $t$ to
represent the domain within the black hole's event horizon as well as
the exterior domain. The integrals in (\ref{eq:lameqn}) and
(\ref{eq:nueqn}) may easily be evaluated to yield the $D$-dimensional
counterpart to Kottler's original solution (a coordinate re-scaling is
assumed):
\begin{eqnarray}
ds^{2}= &-&\left(1-\frac{2M}{r^{D-3}}- \frac{2\Lambda
r^{2}}{(D-1)(D-2)} \right)\, dt^{2} +
\frac{dr^{2}}{\left(1-\frac{2M}{r^{D-3}}-
\frac{2\Lambda r^{2}}{(D-1)(D-2)} \right)} \nonumber \\
&+& r^{2}\,d\Omega^{2}_{(D-2)} \label{eq:dkottler} .
\end{eqnarray}
The term $M/r^{D-3}$ is a potential or harmonic function in a $D-1$
dimensional Euclidean space which is related to the higher dimensional
Newtonian potential.

\subsection{Homogeneous, incompressible star}
\qquad We next study the arbitrary dimensional analogue of the static,
constant density star \cite{ref:schwconst} as well as derive the
corresponding maximum mass/radius relationship. The constant density
sphere is of interest since it will yield the upper limit to the
surface gravitational red shift for spherical (non black hole) systems
in {\em any} dimension. Interesting examples of $2+1$ and $4+1$
dimensional stellar models may be found in
\cite{ref:shir1}$^{-}$\cite{ref:shir3}.

\qquad For the constant density homogeneous sphere we have
\begin{equation}
-T^{t}_{\;t}:=\left\{
\begin{array}{lll}
\rho_{0} & \mbox{ for }  & r < a \\
0 & \mbox{ for }& r > a
\end{array}
\right.
\end{equation}
with $\rho_{0}$ and $a$ some positive constants. All principal
pressures are equal and will be denoted by $p$.

\qquad The only non-trivial conservation equation in the static case
is (\ref{eq:cons2}) from which the condition of hydrostatic equilibrium
may be derived:
\begin{equation}
\frac{1}{2}\nu_{,r}\left(p+\rho\right)+p_{,r}=0. \label{eq:hydrostat}
\end{equation}
The pressure is to be derived from this equation. The metric function
$\lambda(r)$ is given directly by (\ref{eq:lameqn})
\begin{equation}
e^{-\lambda(r)}=1-\frac{16\pi\rho_{0}r^{2}}{(D-2)(D-1)} =: 1-qr^{2},
\label{eq:schwlameqn}
\end{equation}
where it can be seen that the mass of the star increases as $r^{D-1}$.

\qquad By using (\ref{eq:derivs}) and the equation of hydrostatic
equilibrium, one obtains:
\begin{equation}
\sigma_{,r}+\lambda(r)_{,r}\frac{\sigma}{2}-\frac{8\pi
r}{D-2}e^{\lambda(r)}=0 ,
\end{equation}
with $\sigma:=(p+\rho_{0})^{-1}$.

\qquad After much calculation and consideration of boundary conditions
at the surface of the star (the pressure must vanish at the stellar
boundary), the above equation may be solved for the pressure:
\begin{equation}
p=\rho_{0}\left[\frac{(D-3)e^{-\lambda(r)/2}-
(D-3)e^{-\lambda(a)/2}}{(D-1)e^{-\lambda(a)/2}-(D-3)e^{-\lambda(r)/2}}
\right] . \label{eq:schwpreseqn}
\end{equation}
As described in the previous section, this is that last piece of
information required to construct the solution. The metric function,
$\nu(r)$, is now computed directly from (\ref{eq:nueqn}):
\begin{equation}
-g_{tt}:=e^{\nu(r)}= \left[\frac{(D-1)e^{-\lambda(a)/2}-(D-3)e^{-
\lambda(r)/2}}{(D-1)e^{-\lambda(a)/2}-(D-3)}\right]^{2} ,
\end{equation}
and total fluid mass of the sphere is simply
\begin{equation}
M=\frac{8\pi\rho_{0}}{(D-2)(D-1)}a^{D-1}. \label{eq:schwmasstot}
\end{equation}
At this point it may be noted that the pressure becomes infinite at
the same point at which the metric function, $g_{tt}$, vanishes. This
happens when
\begin{equation}
r^{2}=\frac{(D-1)^{2}Ma^{2}-2(D-2)a^{D-1}}{M(D-3)^2}.
\label{eq:infpress}
\end{equation}
To ensure that this does not occur for any acceptable value of $r$, we
must enforce:
\begin{equation}
\frac{M}{a^{D-3}}< \frac{2(D-2)}{(D-1)^2}. \label{eq:condition}
\end{equation}
This is the upper bound on the gravitational potential of the
$D$-dimensional star and is the generalisation of the $M < \frac{4}{9}
a$ law (Buchdahl's theorem \cite{ref:buchdahl}) in four dimensions
which applies to any spherical, static stellar model.

Finally, the metric at the boundary takes on the form:
\begin{equation}
ds^{2}_{|r=a}=-4\frac{e^{-\lambda(a)}}{\left[(D-1)e^{-\lambda(a)/2}-(D-3)\right]^{2}}
\,dt^{2} + e^{\lambda(a)}\,dr^{2} + r^{2}\,d\Omega^{2}_{(D-2)},
\label{eq:schwbmetric}
\end{equation}
from which it can be seen that a coordinate re-scaling:
\begin{equation}
\hat{t}=2\left[(D-1)e^{-\lambda(a)/2}-(D-3)\right]^{-1}t\, ,
\end{equation}
allows smooth joining of (\ref{eq:schwbmetric}) to the $\Lambda\equiv
0$ case of (\ref{eq:dkottler}).

\subsection{Anisotropic fluid}
\qquad Another example is that of the anisotropic fluid star
characterized by:
\begin{eqnarray}
T_{\mu\nu}=(\rho + p_{\perp})u_{\mu}u_{\nu}+
p_{\shortparallel}g_{\mu\nu}+
(p_{\shortparallel}-p_{\perp})s_{\mu}s_{\nu}, \label{eq:anisotmunu} \\
u^{\alpha}u_{\alpha}\equiv -1,\;\;s^{\alpha}s_{\alpha}\equiv +1,\;\;
u^{\alpha}s_{\alpha}\equiv 0. \nonumber
\end{eqnarray}
In the static case, the metric (\ref{eq:spheremetric}) and
(\ref{eq:anisotmunu}) yield
\begin{eqnarray}
&ds^{2}=-e^{\nu(r)}\,dt^{2}+e^{\lambda(r)}\,dr^{2}+
r^{2}\,d\Omega_{(D-2)}^{2}, \label{eq:statmet} \\
u^{t}=e^{-\nu(r)/2},&u^{r}=u^{\theta_{(A)}}\equiv
0,\;\;s^{r}=e^{-\lambda(r)/2},\;\;s^{t}=s^{\theta_{(A)}}\equiv 0.
\nonumber
\end{eqnarray}
Now, the static versions of equations (\ref{eq:lameqn}),
(\ref{eq:nueqn}) and (\ref{eq:latpres}) provide the general solution
of the problem as:
\begin{subequations}
\begin{eqnarray}
e^{-\lambda(r)}=&1-\frac{16\pi}{(D-2)r^{D-3}}\int_{r_{0}}^{r}\rho(x)x^{D-2}\,dx+
\frac{k_{0}}{r^{D-3}}=:1-\frac{2m(r)}{r^{D-3}}, \\
e^{\nu(r)}=&\left[1-\frac{2m(r)}{r^{D-3}}\right]
\exp\left\{c+\frac{16\pi}{D-2}
\int_{r_{0}}^{r}\left[\frac{p_{\shortparallel}(x)+\rho(x)}{x^{D-3}-2m(x)}\right]
x^{D-2}dx\right\}, \\
T^{\theta_{(A)}}_{\theta_{(A)}}:=&\frac{r}{D-2}\left\{p_{\shortparallel
,r}+\left[\frac{1}{2}\nu(r)_{,r}+\frac{D-2}{r}\right]p_{\shortparallel}
+\frac{1}{2}\nu(r)_{,r}\rho(r)\right\}=:p_{\perp}.
\end{eqnarray}
\end{subequations}
Here, $k_{0}$ and $c$ are two arbitrary constants of integration. The
constant $c$ can be absorbed by the transformation $\hat{t}=e^{c/2}t$
and the constant $k$ can be set to zero to avoid a singularity at the
center.

\subsection{D-dimensional Neutron star}
\qquad The neutron star represents a possible end state of a collapsed
massive star. The electrons in the matter making up the star are
subject to extreme pressures and, via the reaction
\begin{equation}
p^{+}+e^{-} \rightarrow n + \nu\;\;, \nonumber
\end{equation}
are converted to neutrons. The bulk of the energy is carried away by
the neutrinos leaving behind a ``cold'' or degenerate remnant.

\qquad The $D$-dimensional uncertainty principle {\em per unit volume}
yields
\begin{equation}
\prod^{D-1}_{j=1}\Delta k^{j} = h^{D-1}, \label{eq:uncprinc}
\end{equation}
where $k^{j}$ is the momentum associated with the $k^{\mbox{th}}$
spatial direction and $h$ is the Planck's constant. By considering a
spherical shell of inner radius $k$ and thickness $\Delta k$ in the
phase space, the number of neutrons per unit volume in this shell will
be the maximum occupation number times the number of cells with phase
volume $h^{D-1}$:
\begin{equation}
N_{n}=\frac{4\pi^{\frac{D-1}{2}}}{\Gamma\left(\frac{D-1}{2}\right)}\frac{k^{D-2}\Delta
k}{h^{D-1}}\; . \label{eq:numneut}
\end{equation}
By integrating (\ref{eq:numneut}) from $k=0$ to the Fermi momentum,
$k_{F}$, we get the maximum number of neutrons per unit volume with
momentum up to $k_{F}$:
\begin{equation}
{\mathcal{N}}=
\frac{4\pi^{\frac{D-1}{2}}\,k_{F}^{D-1}}{\Gamma\left(\frac{D-1}{2}\right)(D-1)h^{D-1}}\;
. \label{eq:numunitvol}
\end{equation}
The energy density of the system is simply given by (using
(\ref{eq:numneut})):
\begin{equation}
\rho=\frac{4\pi^{\frac{D-1}{2}}}{\Gamma\left(\frac{D-1}{2}\right)h^{D-1}}
\int_{0}^{k_{F}}k^{D-2}(k^{2}+m)^{1/2}\,dk , \nonumber
\end{equation}
which, in the extremely relativistic limit (acceptable for interior
regions of neutron stars \cite{ref:MTW}), yields:
\begin{equation}
\rho=\frac{4\pi^{\frac{D-1}{2}}}{\Gamma\left(\frac{D-1}{2}\right)h^{D-1}}
\frac{k_{F}^{D}}{D}\;. \label{eq:neutdens}
\end{equation}

\qquad The pressure of the system is given by differentiating the
energy with respect to volume:
\begin{equation}
p=-\frac{d E}{d
V}=-\frac{d\left(\rho/\mathcal{N}\right)}{d\left(1/\mathcal{N}\right)}
=
\frac{4\pi^{\frac{D-1}{2}}k^{D}}{D(D-1)\Gamma\left(\frac{D-1}{2}\right)h^{D-1}}
=\frac{\rho}{D-1}\; . \label{eq:eqofst}
\end{equation}
This gives the equation of state. Using (\ref{eq:eqofst}) in the
conservation law (\ref{eq:cons2}) and utilising (\ref{eq:derivs})
yields the equation of hydrostatic equilibrium for a $D$-dimensional
neutron star (or any highly relativistic, degenerate $D$-dimensional
Fermi gas):
\begin{equation}
\rho_{,r}=\frac{-D(D-3)\,\rho}{r^{D-2}}\left[1-\frac{2m(r)}{r^{D-3}}\right]^{-1}
\left[\frac{8\pi\rho\,r^{D-1}}{(D-3)(D-2)(D-1)}+m(r)\right].
\label{eq:neuthydrostat}
\end{equation}
The densities in the above equation may be eliminated in favour of
mass terms, giving an O.D.E. for the mass. By assuming a series
solution for $m(r)$, one finds that only one term in the series can
contribute to the solution yielding:
\begin{subequations}
\begin{eqnarray}
m(r)=&\frac{2(D-1)}{D^{2}(D-3)+4(D-1)}r^{D-3}, \label{eq:neutmass} \\
\rho(r)=&\frac{(D-3)(D-2)(D-1)}{4\pi\left[D^{2}(D-3)+4(D-1)\right]}r^{-2}.
\label{eq:neutdens2}
\end{eqnarray}
\end{subequations}
This second equation indicates that the fall-off properties of the
density for the degenerate Fermi gas is {\em independent of
dimension}. The metric functions may now be calculated via
(\ref{eq:lameqn}) and (\ref{eq:nueqn}):
\begin{subequations}
\begin{eqnarray}
e^{-\lambda(r)}=&1-\frac{4(D-1)}{D^{2}(D-3)+4(D-1)},
\label{eq:neutlambda} \\
e^{\nu(r)}=&\left[1-\frac{4(D-1)}{D^{2}(D-3)+4(D-1)}\right]
\left(r/r_{0}\right)^{\frac{4}{D}}. \label{eq:neutnu}
\end{eqnarray}
\end{subequations}
Note that for all $D>3$, the metric given by (\ref{eq:neutlambda}) and
(\ref{eq:neutnu}) is an $R$-domain type metric.

\qquad The above metric derivation is only valid for inner layers of
neutron stars where one can make the extreme relativistic
approximation. The singularity in the density at the center of the
star is also a manifestation due to this approximation. This
singularity plagues the known four dimensional model as well where the
region near $r=0$ is usually excised.

\section{The existence of Birkhoff's theorems}
\qquad One very interesting aspect of spherical symmetry is the
existence of Birkhoff's theorem \cite{ref:birkhoff}. The original
theorem applies to four dimensional systems and states that
spherically symmetric vacua are static and are locally equivalent to
the Schwarzschild solution. This theorem has since been generalised to
include certain matter fields in 4 dimensions including electromagnetic
\cite{ref:hoffmann},\cite{ref:dasem} and scalar fields
\cite{ref:dasscalar},\cite{ref:birkscalar}. A study of the
differentiability properties required for a well posed staticity
theorem may be found in \cite{ref:bergman}. Also, some elegant
techniques have been employed in the literature regarding higher
dimensional versions of the theorem \cite{ref:schmidtbirk},
\cite{ref:bronik}. We briefly present here the conditions required
for, as well as a proof of, a $D$-dimensional Birkhoff's theorem which
applies to both vacuum and non-vacuum systems. The theorem presented
here generalises previous theorems and below we will extend the
theorem to cases where the metric is $C^{0}_{p}$ (piece-wise $C^{0}$).
The assumptions required for the general theorem to hold are minimal
and, most likely, cannot be relaxed, allowing for a most general
theorem. We do, however, examine specific cases where conditions may
be weakened and the theorem will still hold. We state the theorem as
follows:
\\
\begin{theorem}
Let $\mathbf{B}\subset {\mathbb{R}}^{2}$ be a convex domain in the
$r-t$ plane. Let spherically symmetric metric functions $g_{rr}>0$ and
$g_{tt}<0$ belong to the class $C^{3}(\mathbf{B})$ and the
stress-energy tensor, $T^{\mu}_{\;\nu}$, belong to the class
$C^{1}(\mathbf{B})$. Moreover, let: i) $T^{t}_{\;r}\equiv 0$ and ii)
$T^{r}_{\; r,t}\equiv 0$. Then, the metric solutions satisfying
(\ref{eq:einstt})-(\ref{eq:einstthetatheta}) must admit an additional
Killing vector.
\end{theorem}
{\em Proof:} By the assumption on the metric functions we can write
$g_{tt}(r,t)=-e^{\nw}$ and $g_{rr}(r,t)=e^{\lam}$ where $\nw$ and
$\lam$ are of class $C^{3}$. The identity $T^{t}_{\; r}\equiv 0$
yields, from the equation (\ref{eq:einsttr}), that
\begin{equation}
\lambda=\lambda(r). \label{eq:lamproof}
\end{equation}
From this, the equation (\ref{eq:einstt}) implies that
\begin{equation}
T^{t}_{\;t}=T^{t}_{\;t}(r). \label{eq:edenseproof}
\end{equation}
Utilising (\ref{eq:lamproof}) in (\ref{eq:einstrr}) yields:
\begin{equation}
\nu(r,t)_{,r}=\frac{2re^{\lambda(r)}}{D-2} \left\{8\pi T^{r}_{\;r}(r)
+\frac{(D-2)(D-3)}{2r^{2}}\left[1-e^{-\lambda(r)}\right] \right\}.
\end{equation}
Therefore, by differentiability with respect to $t$, we obtain
\begin{equation}
\nu(r,t)_{,r,t} \equiv 0\;\;.
\end{equation}
Therefore, in a convex domain $\mathbf{B}$:
\begin{equation}
\nu(r,t) = \alpha(r) + \beta(t), \label{eq:abproof}
\end{equation}
where $\alpha(r)$ and $\beta(t)$ are differentiable functions.

\qquad Using (\ref{eq:lamproof}) and (\ref{eq:abproof}), the metric
(\ref{eq:spheremetric}) becomes:
\begin{equation}
ds^{2}=-e^{\alpha(r)}e^{\beta(t)}\,dt^{2} + e^{\lambda(r)}\, dr^{2} +
r^{2}\, d\Omega^{2}_{(D-2)} . \label{eq:metricproof}
\end{equation}
By a coordinate transformation,
\begin{equation}
\hat{t}=\int e^{\beta(t)/2}\,dt , \label{eq:transform}
\end{equation}
the metric in (\ref{eq:metricproof}) reduces to a static one which
admits the additional Killing vector
$\frac{\partial}{\partial\hat{t}}\;$. $\blacksquare$
\newline
Note that the only physical assumptions involved in the theorem are
staticity of the radial pressure and that the system possess no
mechanism for radial energy transport. In some circumstances, the
above assumptions may be weakened as will be discussed below.

\qquad Convexity of the domain is required to address certain
situations where the theorem fails though it should seemingly
otherwise hold. Consider the following simple counter-example
\cite{ref:counterbook} depicted in figure 1.
\begin{figure}[ht!]
\begin{center}
\includegraphics[bb=116 117 441 608, clip, scale=0.4,
keepaspectratio=true]{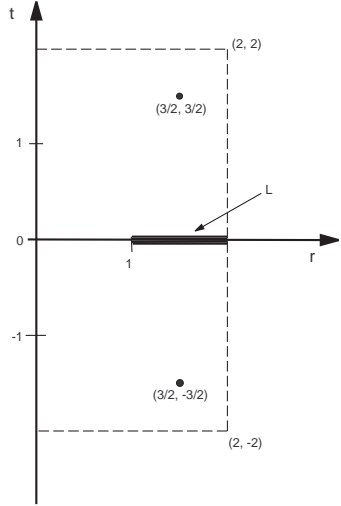} \caption{{\small The non-convex
domain provided by considering the area bounded by the the dashed
line. The line $L$ is removed from the region creating a non-convex
domain. In this case, a staticity theorem will not hold.}} \label{fig1}
\end{center}
\end{figure}

\qquad The line segment, $L$, is given by $L:=\{(r,t) \in
{\mathbb{R}}^{2}:\; 1\leq r < 2;\;\; t\equiv 0\}$. The non-convex
domain is furnished by considering the bounded rectangular domain with
the line, $L$, removed: $\mathbf{B}:= (0,2) \times (-2,2) - L$.
Consider now the metric function, $\lam$, given by
\begin{equation}
\lam=\left\{
\begin{array}{lll}
\frac{1}{5}(r-1)^{5} & \mbox{ for }  & 2 > r > 1,\;\;t > 0  \\
0 & \mbox{ otherwise} &,
\end{array}
\right.
\end{equation}
a $C^{5}$ function with derivatives:
\begin{equation}
\lam_{,r}=\left\{
\begin{array}{lll}
(r-1)^{4} & \mbox{ for }  & 2 > r > 1,\;\;t > 0  \\
0 & \mbox{ otherwise }&
\end{array}
\right.
\end{equation}
and
\begin{equation}
\lam_{,t}\equiv 0. \nonumber
\end{equation}

\qquad Consider the two points $\left(\frac{3}{2},\frac{3}{2}\right)$
and $\left(\frac{3}{2},\frac{-3}{2}\right)$ in the $r-t$ plane. The
function $\lam$ possesses values:
\begin{subequations}
\begin{eqnarray}
\lambda\left(\frac{3}{2},\;\frac{3}{2}\right)&= \frac{1}{160}\;, \nonumber \\
\lambda\left(\frac{3}{2},\;-\frac{3}{2}\right)&= 0. \nonumber
\end{eqnarray}
\end{subequations}
Therefore, although $\lam_{,t}\equiv 0$ in the domain, $\lam$ is {\em
not} independent of time and the metric is non-static. Such arguments
may be applied to cases where non-trivial topological features in the
manifold will create non-convexity in the $r-t$ domain.

\qquad The conditions under which the original, four dimensional
vacuum, theorem hold have been weakened to the point of admitting
a $C^{0}$ metric \cite{ref:bergman} if the metric possesses a
separable $g_{tt}$. We weaken further the conditions here and
demonstrate that a $D$-dimensional metric (not necessarily
vacuum) with separable $g_{tt}$ may possess behaviour as
pathological as {\em piece-wise} $C^{0}$ (a gravitational shock
front) and still be static.

\qquad With little loss of generality, the metric under consideration
may, in the vicinity of the ``jump'', be written as
\begin{equation}
ds^{2}=-e^{\alpha(r)}B^{2}(t)\,dt^{2}+ e^{\lambda(r)}\, dr^{2}
+r^{2}\,d\Omega^{2}_{(D-2)} , \label{eq:discmet}
\end{equation}
where
\begin{equation}
B(t):=2 + \mbox{Sgn} (t)\,. \nonumber
\end{equation}
The function $\mbox{Sgn}(t)$ is given by
\begin{equation}
\mbox{Sgn}(t):=\left\{
\begin{array}{lll}
t/|t| & \mbox{ for }  & t \neq 0 \\
0 & \mbox{ for }& t=0.
\end{array}
\right.
\end{equation}

One may believe that such a shock front would create singular
structure in the manifold since
\begin{equation}
\frac{d\,B(t)}{dt}=2\,\delta(t), \nonumber
\end{equation}
as well as
\begin{eqnarray}
\Gamma^{t}_{tt}&=&\frac{1}{2} \ln\left(-g_{tt}\right)_{,t}\;.
\nonumber \\
\Gamma^{r}_{tt}&=&\frac{1}{2}\alpha(r)_{,r}e^{\alpha(r)-\lambda(r)}
B^{2}(t). \nonumber
\end{eqnarray}
Surprisingly, {\em no} singularity is present in the curvature tensor
and such a solution possesses well behaved orthonormal Einstein tensor.
This is because all derivatives of $g_{tt}$ are multiplied by
quantities which vanish at $t=0$ and it has been rigorously proved
that $0\cdot\delta(t) \equiv 0$ \cite{ref:dasdelta}. The transformation
in (\ref{eq:transform}) is equally well defined as:
\begin{eqnarray}
\hat{t}&=&\int_{t_{0}}^{t} B(\tau)\,d\tau \,\;\;\;\;t_{0}<0 \nonumber
\\
&=&\left\{
\begin{array}{lll}
t+|t_{0}| & \mbox{ for } & t_{0} \leq 0 \\
3t+|t_{0}| & \mbox{ for }&  t > 0.
\end{array}
\right.
\end{eqnarray}
Notice that the re-scaled time variable, $\hat{t}$, is continuous at
at $t=0$. Such a transformation is therefore admissible and the
re-scaled metric is explicitly static. It is unknown if behaviour
worse than $C^{0}_{p}$ (such as characteristic fluctuations) may admit
a staticity theorem.

\qquad It is well known that hyperbolic equations admit exact
discontinuous solutions and that Einstein's equations therefore
allow for such solutions. Before $t=t_{0}$ the metric under
consideration can be transformed smoothly to a static one. After
$t=t_{0}$ a similar transformation also renders a static metric.
At $t=t_{0}$ there exists a $C^{0}$ transformation to a static
metric. Therefore, in the entire domain, the metric can be
transformed to a static one although the transformation is not
the usual $C^{3}$.

\qquad The discontinuous metric in the form of (\ref{eq:discmet})
yields discontinuous orthonormal tetrad as well as the singular
Christoffel symbols mentioned above. However, the {\em Riemann
invariants}, which govern the gravitational physics possess a
removable dicontinuity which may be ignored. The potentially
problematic components are:
\begin{eqnarray}
R_{\hat{t}\hat{r}\hat{t}\hat{r}}&=&\frac{1}{2}
e^{-\lambda(r)}\left[ \alpha(r)_{,r,r}+\frac{1}{2}\alpha_{,r}^{2}
-\frac{1}{4}\alpha(r)_{,r}\lambda(r)_{,r}\right], \nonumber \\
R_{\hat{t}\hat{\theta}\hat{t}\hat{\theta}}&=&\frac{1}{2r}\alpha(r)_{,r}e^{-\lambda(r)},
\nonumber
\end{eqnarray}
where hatted indices denote quantities calculated in the
orthonormal frame. Notice that the Riemann invariants are well
behaved and \emph{independent} of the function $B(t)$. These
components are all static indicating staticity of the
gravitational field and therefore yielding a Birkhoff's theorem.

\qquad Generally, a $C_{0}^{p}$ metric yields serious
singularities in the manifold (see \cite{ref:aichsexl},
\cite{ref:sfetsos} for studies of such systems). However, for the
type of metric given by (\ref{eq:discmet}), singularities are
avoided.

\qquad Finally, we should mention that results in this section would
be unaltered if one relaxes the condition for the space-time to
possess a strictly Lorentzian metric. In such case, the metric
(\ref{eq:metricproof}) is better written as
\begin{equation}
ds^{2}=-e^{\alpha(r)}A(t)\,dt^{2} + e^{\lambda(r)}\,dr^{2} +
r^{2}\,d\Omega^{2}_{(D-2)},
\end{equation}
where the function $A(t)$ may switch sign at one or more values of $t$,
such as $A(t)=t^{3}$ (see \cite{ref:dascollapse} \cite{ref:tdomain}
for examples of such space-times). The Birkhoff's theorem holds both
in the Lorentzian branch and the Euclidean branch of the manifold.
Also, in the case where the signs of $g_{tt}$ and $g_{rr}$ are both
switched, we obtain the generalisation of Birkhoff's theorem in the
$T$-domain \cite{ref:hawkellis}. In this situation, the theorem is a
homogeneity theorem as opposed to a staticity theorem since the
additional killing vector is space-like.

\section{Concluding remarks}
\qquad We have considered spherically symmetric solutions to the
Einstein field equations with an arbitrary number of spatial
dimensions. A reasonably general method has been presented which
allows one to solve, at least in quadrature, these equations. The
method has been illustrated by quickly and efficiently computing the
metric for a $D$-dimensional black hole with arbitrary cosmological
constant, the metric for a $D$-dimensional constant density star,
anisotropic fluid star and neutron star. An upper limit has been
placed on the mass/radius ratio of stars of arbitrary dimension.
Finally, the minimum general requirements for a $D$-dimensional
Birkhoff's theorem in both vacuum and non-vacuum systems has been
presented. To have a rigorous theorem, one must insist on convexity of
the domain in question. In certain situations, a staticity theorem may
hold even when the metric component, $g_{tt}$, is only $C^{0}_{p}$ or
piece-wise continuous. It may be shown that the theorem also holds for
metrics of Euclidean signature as well as inside black hole
$T$-domains.
\newpage
\bibliographystyle{unsrt}





\end{document}